# Early Mars volcanic sulfur storage in the cryosphere and formation of transient $SO_2$-rich atmospheres during the Hesperian

F. SCHMIDT[1,2], E. CHASSEFIÈRE[1,2], F. TIAN[3], E. DARTOIS[4], J.-M. HERRI[5], AND O. MOUSIS[6]

[1]Université Paris Sud; GEOPS, UMR 8148, Bât. 504, Orsay F-91405, France
[2]CNRS; Orsay F-91405, France
[3]Center for Earth System Sciences, Tsinghua University, Beijing, China
[4]IAS, Université Paris-Sud, CNRS, France
[5]Centre SPIN, ENS des Mines of Saint-Etienne, France
[6] Aix Marseille Université, CNRS, LAM (Laboratoire d'Astrophysique de Marseille) UMR 7326, 13388 Marseille, France

**Abstract**- In a previous paper (Chassefière et al., *Icarus* 223, 878-891, 2013), we have shown that most volcanic sulfur released to early Mars atmosphere could have been trapped in the cryosphere under the form of $CO_2$-$SO_2$ clathrates. Huge amounts of sulfur, up to the equivalent of a ~1 bar atmosphere of $SO_2$, would have been stored in the Noachian cryosphere, then massively released to the atmosphere during Hesperian due to rapidly decreasing $CO_2$ pressure. It would have resulted in the formation of the large sulfate deposits observed mainly in Hesperian terrains, whereas no or little sulfates are found at the Noachian. In the present paper, we first clarify some aspects of our previous work. We discuss the possibility of a smaller cooling effect of sulfur particles, or even of a net warming effect. We point out the fact that $CO_2$-$SO_2$ clathrates formed through a progressive enrichment of a preexisting reservoir of $CO_2$ clathrates and discuss processes potentially involved in the slow formation of a $SO_2$-rich upper cryosphere. We show that episodes of sudden destabilization at the Hesperian may generate 1000 ppmv of $SO_2$ in the atmosphere and contribute to maintaining the surface temperature above the water freezing point.

# INTRODUCTION

For a long time it has been thought that a dense $CO_2$-dominant atmosphere was necessary in order to keep early Mars warm and wet. However, the most recent 3D models of the early Martian climate considering the possibility of several bar $CO_2$ atmospheres fail in producing surface temperature higher than the freezing point of water (Forget et al. 2013; Wordsworth et al. 2013). It has been recently shown that an $H_2$-rich (10-20% volume mixing ratio) 1-2 bar $CO_2$ atmosphere may heat the early Mars surface above the freezing point of water (Ramirez et al. 2014). The source of $H_2$ could be the mantle, or a $CH_4$-enriched cryosphere following serpentinization (Chassefière et al., this issue).

Sulfur dioxide ($SO_2$) has been proposed as a possible greenhouse gas working together with $CO_2$ to raise the surface temperature of early Mars above the freezing point of water (Halevy et al. 2007; Johnson et al. 2008). However the cooling effect of pure $H_2SO_4$ (and $S_8$) particles formed in the atmosphere might counteract the warming due to $SO_2$ greenhouse effect. After a short period of warming, the presence of sulfate aerosols would have resulted in a colder surface, with a net cooling, instead of warming, of the planet (Tian et al. 2010). The cooling effect of sulfur could have resulted in the trapping of most of the volcanically released $SO_2$ in the cryosphere under the form of $CO_2$-$SO_2$ clathrates (Chassefière et al. 2013, this reference is denoted by C2013 in the following). If so, the early Mars cryosphere could have accumulated huge amounts of frozen sulfur, later released to the atmosphere, which would be at the origin of the large sulfate deposits observed on Hesperian terrains (C2013).

It has been argued that $H_2SO_4$ particles are not pure, but mixed with pre-existing suspended aerosols in a dusty early atmosphere and that a net warming can occur for a $SO_2$ mixing ratio larger than 1 ppmv (Halevy and Head 2014a). This warming might be substantial in tropical regions due to punctuated intense volcanic emissions. Such warm episodes would not last more than a tens or hundred years, even after exceptionally strong volcanic eruptions. In the case of pure $H_2SO_4$ crystals (no dust in the atmosphere), the net cooling effect due to the formation of sulfur particles in early Mars average atmosphere is of the order from 20 K (at 1 ppmv $SO_2$) to 60 K (at 10 ppmv $SO_2$) (Tian et al. 2010). For an average dusty atmosphere, particles are composed of a dust nucleus and a mantle of condensed $H_2SO_4$, with reflective particles, and the net effect is a warming in the range from 15 K (at 1 ppmv $SO_2$) to 20 K (at 10 ppmv $SO_2$) in comparison to the already cooled dusty environment (Halevy and Head, 2014a). It is noticed that Halevy and Head (2014a) did not included the S particle, which was included in Tian et al., 2010. There is therefore an uncertainty on the true net radiative effect, which is dependent on the assumed structure of particles and the dust load.

Sulfur, under both oxidized ($SO_2$) and reduced ($H_2S$) forms, could have been released in large amounts by volcanism during the Noachian. From several 100 mbar to ~1 bar of sulfur could have been outgassed along Martian history, most of which during the Noachian and the Hesperian (e.g., Craddock and Greeley 2009; Gaillard and Scaillet 2009). 1 bar of sulfur corresponds to a mass of sulfur of ~4 × $10^6$ Gt. For comparison, an exceptionally strong ten-year-long eruption at 1000 times the terrestrial average sulfur outgassing rate would inject ~25 Gt of sulfur in the Martian atmosphere (Halevy and Head 2014b). The total amount of sulfur outgassed by Mars over the whole Noachian and the beginning of the Hesperian represents typically ~100,000 times the sulfur ejected by one such intense eruption. According to Halevy and Head (2014a) a series of typically 100,000 large eruptions every ≈1000 years during 100-200 Myr, with peak atmospheric $SO_2$ mixing ratios of 1-2 ppmv during a few ten years, would have been sufficient to allow liquid water to carve valley networks along a series of sporadic warming events, with a total duration of warm conditions of several million years (Hoke et al. 2011). Alternatively, a small or moderate number of late global destabilization events of a sulfur-rich early Mars cryosphere, with much higher levels of $SO_2$ in the atmosphere (up to tens to hundred ppmv) may have occurred during the Hesperian (C2013). According to this scenario, the long-term storage of sulfur in the cryosphere, followed by a massive release to the atmosphere, would be at the origin of the massive sulfate deposition that occurred at the Noachian (C2013).

The present article aims at estimating the consequence on clathrate formation and sulfur trapping in the scope of aerosol radiative effect. First, we recall the main results of our previous study (C2013), that is the potential trapping of most sulfur volcanically outgassed during the first billion years in the cryosphere under the form of $CO_2$-$SO_2$ clathrates, and its release to the atmosphere at the Hesperian due to rapid atmospheric pressure decrease. This is the case of the net cooling hypothesis, relevant if the early Mars atmospheric dust load was low. In the next section, the consequence of net warming due to dusty aerosol (Halevy and Head 2014b) on clathrate formation will be discussed for completeness purpose. In the following sections, we clarify the impact of a possibly smaller cooling effect of sulfur particles. We also discuss physical mechanisms potentially involved in the progressive enrichment of $CO_2$ clathrates with $SO_2$ during the slow formation of the sulfur reservoir during the Noachian. We show that episodes of sudden destabilization at the Hesperian may generate 1000 ppmv of $SO_2$ in the atmosphere, which could have helped contribute to maintain the surface temperature above the water freezing point. In the conclusion section, we assess the potential role of $SO_2$ in contributing to warming episodes responsible for the formation of valley networks.

## 1. ALWAYS NET COOLING: EARLY MARS VOLCANIC SULFUR STORAGE IN THE CRYOSPHERE

Because the background early Mars climate has been likely cold and dry, it has been hypothesized that virtually all volcanic sulfur released during the Noachian has been trapped in the superficial cryosphere under the form of $CO_2$-$SO_2$ clathrates (C2013). Using a one-dimensional radiative chemical model (from Tian et al. 2010), we have shown that, above a threshold of 2 bar of $CO_2$, any additional $CO_2$ released by volcanism condenses as $CO_2$-clathrates, either directly at the surface, or through condensation into atmospheric clathrate particles which further precipitate. The reason for this 2 bar limit is simple. For increasing $CO_2$ pressure above 2 bars, the albedo of the planet increases due to more efficient Rayleigh scattering and the surface temperature drops below the $CO_2$ clathrate thermodynamic equilibrium temperature (cf Fig. 2 in C2013). Assuming a 2-bar atmosphere, that is at saturation with respect to $CO_2$ clathrate formation, all volcanically released $SO_2$ and the fraction of atmospheric $CO_2$ in excess of 2 bars is converted into $CO_2$-$SO_2$ clathrates soon after release. The reason why virtually all $SO_2$ is converted is that $SO_2$ forms clathrates much more easily than $CO_2$, with an increase of the $SO_2/CO_2$ ratio by 2 orders of magnitude in the clathrate with respect to the gas phase (cf Fig. 3 in C2013, keeping in mind that $SO_2/CO_2$ still stays <<1).

The threshold pressure value of 2 bars has been obtained by using the 1-D model. The mean an-nual surface temperature is not uniform, decreasing from the equator to the poles. By analogy with present Mars, it appears likely that even with a $CO_2$ pressure of only 0.5-1 bar, regions above 45° latitude could have been below the clathrate formation temperature in annual average (CA). If so, the cold trap formed by middle and high latitude regions may have been efficient enough to remove most of the $SO_2$ released to the atmosphere (including in equatorial regions) by volcanism. More accurate 2-D or 3-D models would be needed to precisely define the value of the pressure threshold, which is smaller than 2 bars. Such a lower threshold value is in better agreement with a strong carbon hydrodynamic escape ruling out a dense Martian atmosphere until the late Noachian (Tian et al., 2009), and various constraints on early Martian atmospheric pressure based on geochemical and geomorphological arguments (e.g., Kite el al. 2014). From all existing constraints, an average $CO_2$ pressure value of ~1.5 bar at the Noachian/ Hesperian transition could be close to the truth (Chassefière et al. this issue).

Assuming that the $CO_2$ pressure sharply decreased during the Hesperian from ~1.5 bar to less than 0.5 bar, huge amount of sulfur would have been outgassed from the cryosphere due to the destabilization of $CO_2$-$SO_2$ clathrates generated by atmospheric depressurization. Although it could have been episodic, up to 4 million Gt of sulfur would have been outgassed in a few ten or hundred millions years. In terms of released amount of sulfur, it would be equivalent to a very large eruption injecting 40 Gt in the atmosphere every 1000 years, or a continuous emission 20 times the terrestrial average sulfur outgassing rate, during 100 Myr. C2013 have suggested that massive emplacement of sulfate minerals during the Hesperian possibly by aerosol precipitation (Michalski et al., 2011), whereas only very tiny deposits are observed on Noachian terrains (Bibring et al., 2006), could be explained by such a long-term storage of virtually all sulfur volcanically outgassed during the Noachian, and abrupt release of this sulfur to the atmosphere during the Hesperian atmospheric depressurization phase.

## 2. ALWAYS NET WARMING: NO SULFUR STORAGE IN THE CRYOSPHERE

If the scenario of a net warming occurred due to a globally dusty atmosphere, the episodic volcanic activity warmed the atmosphere thanks to $H_2SO_4$-bearing aerosols (Halevy and Head 2014a). Afterward, such warming episode declined by aerosols precipitation in a wet environ-ment at a temperature much above the clathrate formation in subsolar region. Nevertheless, the mean temperature is close to the clathrate stability (215K at 1 bar) so that clathrate may form in the cold polar and high altitude regions.

In this case, the previous discussion of a 2-bar buffered atmosphere is not valid anymore. In the first step of the cycle, when the warming occurs, the $SO_2$ and $H_2SO_4$-bearing aerosols will at least partly compensate the cooling due to both Rayleigh scattering of the $CO_2$ and Mie scattering of the dust particles. If the temperature near the equator is high enough, a wet regime could occur with positive feedback: the atmosphere is dried by precipitation of $H_2SO_4$ and the temperature increases again due to $SO_2$. In the second step, the volcanic $SO_2$ decreases again, the temperature goes back to normal and the $CO_2$-$SO_2$ clathrate will condensate again. The few $SO_2$ still present in the atmosphere will condensate preferentially into clathrate as explained in C2013. Finally, the system converges to the 2-bar buffered atmosphere before another volcanic cycle. By increasing the temperature, the surface pressure in the stability of clathrate is also enhanced (3 bar at 240K, 10 bars at 265K) but in this scenario, the high temperatures are only reached in equatorial region. Thus clathrate will still form in the cold polar regions, as the average temperature is always cold (200K at 1 ppmv, from Halevy and Head 2014a). The competition between $SO_2$ removal by wet aerosol scavenging at high latitude and $CO_2$-$SO_2$ clathrate condensation at high latitude is not easy to evaluate and needs further modeling.

## 3. FORMATION OF TRANSIENT $SO_2$-RICH ATMOSPHERES AT THE HESPERIAN

### 3.1 The effect of a weaker cooling by sulfur aerosols

C2013 uses the radiative transfer model of Tian et al. (2010) assuming pure $H_2SO_4$ aerosol particles and obtains a strong cooling due to the albedo feedback. It has been proposed that mixed dust-$H_2SO_4$ particles would be less efficient in cooling the atmosphere, although an early Mars with a dusty atmosphere would have been a colder place in comparison with the same planet with a clear sky before the release of $SO_2$ into the atmosphere (Halevy et al. 2014a). What would be the effect of a weaker cooling on the C2013 scenario?

It is assumed in C2013 that atmospheric $CO_2$ is at saturation with respect to $CO_2$ clathrate formation (2 bar in a one–dimensional approach). Because $SO_2$ in the $CO_2$ clathrates is enriched by a factor of ~100 with respect to $SO_2$ in the gas phase, virtually all volcanically released $SO_2$ is stored in the $CO_2$ clathrates once the later is formed. The equilibrium $SO_2$ level in the atmosphere is radiatively controlled by the condition that the surface temperature being equal to the $CO_2$ clathrate formation temperature. Due to the cooling effect, the equilibrium $SO_2$ level is 30 ppmv and 200 ppmv at 1bar/215 K and 0.5 bar/205K respectively. Below 0.5 bar, there is no more equilibrium value and all the $SO_2$ released by volcanoes may remain in the atmosphere. Thus in the scenario with a strong radiative cooling effect from sulfate aerosols, $SO_2$ released by volcanoes may remain at a progressively higher level in the atmosphere after the eruption when the pressure decreases below the saturation value of 2 bars - trapping of sulfur in clathrates becomes less and less efficient for decreasing pressure.

First let's consider the case with no $SO_2$ radiative cooling or warming. When the atmospheric $CO_2$ level is at saturation with respect to condensation into $CO_2$ clathrates, the amount of $SO_2$ remaining in the atmosphere after an eruption is close to zero. Indeed, any volcanic release of $CO_2$ results in the trapping of an equivalent quantity of $CO_2$ in the cryosphere. As soon as the $CO_2$ pressure decreases below its saturation value, all $SO_2$ released by volcanoes remains in the atmosphere and there is no more trapping of $SO_2$ in clathrates. A comparison between the two cases is displayed in Figure 1. Assuming that cooling is weaker than assumed in C2013, the situation is intermediate between the two dashed curves in Fig. 1.

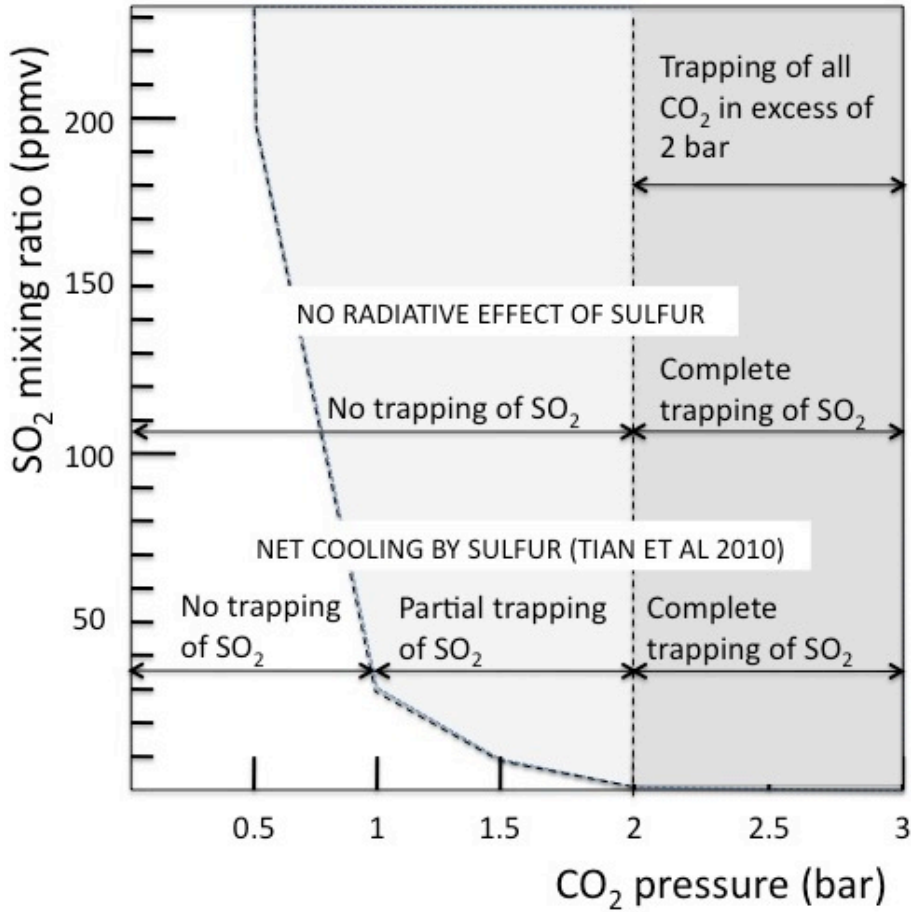

*Figure 1: In the case of a net cooling due to sulfur aerosols (Tian et al. 2010), SO₂ released by volcanoes may remain at a progressively larger level in the atmosphere after the eruption when the pressure decreases below the saturation value of 2 bars. Trapping of sulfur in clathrates is less and less efficient for decreasing pressure. Due to the cooling effect of sulfur, the equilibrium SO₂ level in the atmosphere is such as the radiatively controlled surface temperature is equal to the CO₂ clathrate formation equilibrium temperature. At 1 bar, this equilibrium value is 30 ppmv, and it increases up to 200 ppmv at 0.5 bar. Below 0.5 bar, there is no more equilibrium value and all the SO₂ released by volcanoes may remain in the atmosphere. If sulfur is neutral, with no radiative effect, all SO₂ released by volcanoes after the time when pressure drops below the saturation level remains in the atmosphere, and there is no more trapping of SO₂ in clathrates.*

**3.2 Positive feedback in an episodically warming dusty atmosphere**

The dust load of the Martian atmosphere may be not constant but episodically high and low. After a period of low dust content, the atmosphere behaves as a net cooling and the cryosphere is trapping $SO_2$ (see section 3.1). In a period of high dust load, the net warming by $SO_2$ release will trigger a strong positive feedback: higher temperature leads to clathrate destabilization and more $SO_2$ release, that again leads to higher temperature. The sulfur trapped in the cryosphere will substantially be released, even a "small" eruption may have a potentially large effect (much larger than modeled by Halevy and Heab, 2014a). Did a significant number of such positive feedback episodes occur on early Mars? It seems unlikely, due to the tiny sulfate deposits observed on Noachian terrains (C2013). The absence of large deposits of sulfate minerals on Noachian terrains might suggest that no significant wide scale positive feedback occurred, pointing toward a net cooling rather than a net warming effect of sulfur at global scale. It could also mean that the early mars atmosphere was not dusty.

**3.3 Progressive enrichment of $CO_2$ clathrates with $SO_2$ during the slow formation of the sulfur reservoir during the Noachian**

It has been implicitly assumed by C2013 that the global $CO_2$-$SO_2$ clathrate reservoir is not at thermodynamic equilibrium with the atmosphere during the long phase of accumulation of $SO_2$ in $CO_2$ clathrates. In the course of freshly released $SO_2$ trapping in $CO_2$ clathrates, there is a progressive enrichment of superficial clathrate layers with $SO_2$. The upper cryosphere is stratified, with new layers of $CO_2$-$SO_2$ clathrates successively deposited at the top, knowing that both $CO_2$ and $SO_2$ (together with $H_2S$, rapidly converted to $SO_2$ by atmospheric photochemical processes) are released during volcanic eruptions. The progressive enrichment of clathrates with $SO_2$ from an initial value of a few hundred ppmv of $SO_2$ at the beginning in the top layer in thermodynamic equilibrium with the atmosphere, up to 10% or more in the whole clathrate reservoir at the end of the Noachian (typically one or a few hundred meters thick), has not been modeled.

Several processes would have to be taken into account in such a modeling. First, the progressive burying of older clathrate under younger overlying layers of freshly formed clathrates results in an increase of the cryostatic pressure over buried layers, increasing their thermodynamic stability and sealing them from the atmosphere. Assuming that at the beginning the substrate is pure $CO_2$ clathrate, the addition of $SO_2$- rich new clathrates at the top will result in a vertical positive gradient of $SO_2$ mixing ratio in the clathrates. A downward diffusion of $SO_2$ through the clathrates is therefore expected. The kinetics of the diffusion of gases in clathrates is poorly known. The measured diffusion coefficients of $CH_4$ and $CO_2$ in clathrates are in the range from $1.0 \times 10^{-15}$ – $3.8 \times 10^{-10}$ $m^2$ $s^{-1}$ (references in Mousis et al. 2012). Using the highest value, the diffusion time over a depth of 100 m is $3~10^{13}$ s (1 Myr). Most of these measurements have been made between 250 and 270 K, 20-50 K above the mean Martian surface temperature, and because diffusion coefficients decrease exponentially with temperature, these time scales are certainly higher limits. The diffusion coefficient of $SO_2$ in clathrates is unknown. For a diffusion coefficient larger than a threshold value of $1.0 \times 10^{-12}$ $m^2$ $s^{-1}$, the homogenization of a gas in clathrates over a typical depth of 100 m is efficient over time scales from a few to a few hundred million years, compatible with the duration of the Noachian period. In the absence of measurements, the efficiency of diffusion in transporting $SO_2$ from the top to the deeper reservoir by diffusion cannot be assessed.

$SO_2$ transport through clathrates can be made easier by cracks in the clathrate lattice, and/or local decomposition/ recrystallization events. Due to their much stronger affinity with the clathrate structure, $SO_2$ molecules, provided they are sufficiently mobile, will tend to replace $CO_2$ molecules and increase the $SO_2/CO_2$ ratio in the clathrate. Due to the lack of knowledge of such microphysical processes, it is not possible to quantitatively assess their efficiencies in transporting $SO_2$ and enriching the clathrate with sulfur.

Another important property of clathrates is their metastability. The phenomenon of "anomalous preservation" of clathrates below the freezing point of water has been experimentally observed since more than 20 years ago for a great variety of clathrate hydrate systems and has potential application for successful retrieval of natural gas hydrates (Stern et al. 2011 and references therein). It has been observed that, at 1 bar, clathrate particles may be stable at temperatures above the thermodynamic equilibrium temperature and preserve most of their structure during rapid release of the pressure of the confining system where they have been synthesized (e.g., Stern et al. 2011 and references in Chassefière 2009). Upon heating through the water freezing point (273 K), the particles rapidly decompose and release all their residual gas. Transposing the situation to the ~1 situation to the ~1 bar $CO_2$ atmosphere and low temperatures on early Mars, anomalous preservation contributes to prevent clathrates from decomposing. Due to both the progressive burying of clathrates under new overlying layers, with an increase of the cryostatic pressure making them more stable, and their metastability, preventing them from decomposing due to small fluctuations of temperature and pressure, it is expected that the clathrate reservoir at the top of the cryosphere is globally cohesive and stable over long periods of time.

**3.4 Consequences of a sudden decomposition of clathrates and resulting thermodynamic equilibrium between clathrate and atmosphere at late Noachian/Hesperian times**

An abrupt decomposition of a $SO_2$-rich upper cryosphere (e.g. $SO_2/CO_2$ = 20%, cf C2013) results in the injection of large amounts of $SO_2$ into the atmosphere. It may occur if the global average surface temperature rises above the freezing point of water, due to a triggering event such as a giant impact or a massive outgassing of hydrogen (Ramirez et al. 2014). If so, anomalous preservation is broken and clathrates decompose to liquid water and release trapped gas molecules to the atmosphere. Outgassing occurs until a thermodynamic equilibrium is reached between the clathrate and the gas phase. At 220 K, the enrichment factor of $SO_2$ with respect to $CO_2$ in the clathrate is in the range from 100-400 with respect to the gas phase (cf Fig. 3 in C2013). Assuming a $SO_2$ mixing ratio of 20% in the clathrate, the gas phase must contain 500-2000 ppmv of $SO_2$ with respect to $CO_2$ to be at equilibrium with the clathrate. It results that the mixing ratio in the atmosphere rapidly increases up to ~1000 ppmv after destabilizing the clathrates. With 1000 ppmv of $SO_2$, at $CO_2$ pressure close to 1 bar, the greenhouse warming effect of $SO_2$ raises the surface temperature up to the freezing point of water (C2013). If so, the greenhouse effect of $SO_2$ can take the relay of the initial triggering event, by maintaining the surface temperature above 0°C and allowing the development of a full hydrological cycle, even when the triggering effect has disappeared.

Without an independent triggering effect, such a heating of the planet by large amounts of $SO_2$ suddenly outgassed from the cryosphere may seem difficult to achieve. If the formation of sulfate particles results in a net cooling effect (Tian et al. 2010), the large greenhouse effect will be rapidly counteracted in the course of the formation of aerosols, typically one month or a few months. Only in the case of a sudden (taking less than a few days) release of the equivalent of 1000 ppmv of $SO_2$ in a background 1-2 bar $CO_2$ atmosphere may occur before cooling is effective, a heating is possible, but this hypothesis is not favored. In the case of a dusty early Mars atmosphere resulting in a net warming effect (Halevy and Head 2014), the triggering of global hydrological episodes by large volcanic eruptions is possible on equatorial region if the heating dominates the cooling due to dust aerosols.

**3.5 Frequency and intensity of catastrophic disruption in Noachian/Hesperian transition**

In the case of a net cooling by sulfate aerosols, the proposed mechanism may be quite efficient in relaying a global warming episode due to another cause (e.g. big impact or $H_2$ release). A temperature rising above the freezing point of water would automatically trigger the destabilization of metastable $SO_2$-rich clathrates, with the release to the atmosphere of the amount of $SO_2$ required to put clathrates in thermodynamic equilibrium with the gas phase. Assuming that the $SO_2$ atmospheric content rapidly climbs up to 1000 ppmv, which corresponds to a plausible $SO_2/CO_2$ ratio of 0.2 in the clathrate reservoir following accumulation of volcanic sulfur during the Noachian (C2013), the amount of released $SO_2$ is only a small part of the content of the global clathrate reservoir. If 1 bar of $SO_2$ has been trapped, only 0.1 % of the reservoir content is necessary to reach 1000 ppmv of $SO_2$. 1000 ppmv nevertheless represent ~4000 Gt of sulfur in the atmosphere, ~2 orders of magnitude above the quantity released by the strongest expected volcanic eruptions on early Mars (Halevy and Head 2014b). A maximum of ~1000 such episodes could have occurred sequentially until full depletion of the clathrate reservoir. More plausibly such an initial warming episode should have resulted in a steady state outgassing/sulfate formation process, with a release rate from the global clathrate reservoir at equilibrium with the atmosphere controlled by the loss rate of sulfur through sulfate deposition. Further modeling is necessary to estimate the duration of such a global event. Such modeling is a necessary step to estimate the capacity of an active $SO_2$ cycle to maintain liquid water at the surface during the time required for the formation of valley networks, typically 0.1-10 Myr (Hoke et al. 2011).

**CONCLUSION**

In this article, we discussed the clathrate formation and the potentially trapped sulfur in the cryosphere in the light of recent photochemical-thermodynamical model. We found that even in the case of a net warming, the cryosphere and clathrate should form at high latitude. The competition between wet equatorial scavenging of $SO_2$ particles and $CO_2$-$SO_2$ clathrate condensation at the pole must be evaluated in future studies.

Surface pressure could be higher than 2 bars only in the case of a net warming of sulfate aerosol to prevent clathrate to condensate due to strong cooling caused by Rayleigh scattering of $CO_2$ gas.

Strong positive feedbacks are expected if the dust load in episodic, leading to intense hot periods. The sudden release of $SO_2$ in a dust rich atmosphere trigger a net warming, whereas it trigger a net cooling in a dust free atmosphere. If the atmosphere is globally dust free, net cooling dominates and the cryosphere should enrich in $SO_2$. If then an episodic period of high dust load occurs, the net warming should destabilize the cryosphere, enriching the atmosphere in $SO_2$, triggering a strong positive feedback.

In the future, more modeling and experimental works are required in order to better quantify the expectation proposed here. In particular, the diffusion of $SO_2$ and $CO_2$ in the solid clathrate phase is required to estimate more precisely the potential cryosphere enrichment. Also the combined effects of $SO_2$, elemental S and sulfate aerosols in a realistic early Mars atmosphere are needed. Finally, the modeling of an active $SO_2$ cycle to maintain liquid water at the surface is necessary to estimate the time scale of such transient period.

**ACKNOWLEDGEMENTS**

Frédéric Schmidt and Eric Chassefière acknowledge support from "Institut National des Sciences de l'Univers" (INSU), the "Centre National de la Recherche Scientifique" (CNRS) and "Centre National d'Etude Spatiale" (CNES) and through the "Programme National de Planétologie" and MEX/PFS Program. Olivier Mousis thanks the support of the A*MIDEX project (n° ANR-11-IDEX-0001-02) funded by the «Investissements d'Avenir » French Government program, managed by the French National Research Agency (ANR). Feng Tian is supported by the National Natural Science Foundation of China (41175039) and the Startup Fund of the Ministry of Education of China.

# REFERENCES

Bibring J.-P., Langevin Y., Mustard J. F., Poulet F., Arvidson R., Gendrin A., Gondet B., Mangold N., Pinet P., and Forget F. 2006. Global Mineralogical and Aqueous Mars History Derived from OMEGA/Mars Express Data. *Science* 312, 5772, 400-404

Chassefière, E., 2009. Metastable methane clathrate particles as a source of methane to the Martian atmosphere, Icarus 204, 137-144.

Chassefière E., Dartois E., Herri J.-M., Tian F., Schmidt F., Mousis O., and Lakhlifi A. 2013. $CO_2$-$SO_2$ clathrate hydrate formation on early Mars. *Icarus* 223, 878-891. (C2013)

Craddock R. A., and Greeley R. 2009. Minimum estimates of the amount and timing of gases released into the martian atmosphere from volcanic eruptions. *Icarus*. 204, 512-526.

Forget F., Wordsworth R., Millour E., Madeleine J.-B., Kerber L., Leconte J., Marcq E., and Haberle R.M. 2013. 3D modelling of the early martian climate under a denser CO2 atmosphere : Temperatures and CO2 ice clouds. Icarus 222, 81-99

Gaillard F., and Scaillet B. 2009. The sulfur content of volcanic gases on Mars. *Earth Planet. Sci. Lett*. 279, 1-2, 34-43.

Gaillard F., Michalski J., Berger G., McLennan S., and Scaillet B. 2012. Geochemical reservoirs and timing of sulphur cycling on Mars. *Space Sci. Rev*. 174, 251-300, doi 10.1007/s11214-012-9947-4.

Halevy I., Zuber M. T., and Schrag D. P.. 2007. A Sulfur Dioxide Climate Feedback on Early Mars. *Science* 318, 5858, 1903-1907.

Halevy I., and Head J.W. 2014a. Episodic warming of early Mars by punctuated volcanism. Nature Geosci. 7, 865-868, doi:10.1038/ngeo2293.

Halevy I., and Head J.W. 2014b. Climatic and chemical consequences of episodic eruptions on early Mars. The Fifth International Workshop on the Mars Atmosphere: Modelling and Observation, held on January 13-16 2014, in Oxford, U.K. Edited by F. Forget and M. Millour, id.4204.

Hoke M.R.T., Hynek B.M., and Tucker G.E. 2011. Formation time scales of large Martian valley networks. *Earth Planet. Sci. Lett.,* 312, 1-12.

Johnson S. S., Mischna M. A., Grove T. L., and Zuber M. T. 2008. Sulfur-induced greenhouse warming on early Mars. *J. Geophys. Res*. 113, E8, CiteID E08005.

Kite E.S., Williams J.-P., Lucas A., and Aharonson O. 2014. Low palaeopressure of the martian atmosphere estimated from the size distribution of ancient craters. *Nat. Geosci.*, 7, 335-338, doi: 10.1038/ngeo2137.

Michalski, J. & Niles, P. B., 2011, Origin of Martian Sulfates and Interior Layered Deposits (ILDs) in the Valles Marineris by atmospherically driven processes, EPSC-DPS Joint Meeting 2011, 1752.

Mousis O., Chassefière E., Lasue J., Chevrier V., Elwood Madden M. E., Lakhlifi A., Lunine J.I., Montmessin F., Picaud S., Schmidt F., and Swindle T.D. 2013. Volatile Trapping in Martian Clathrates. *Space. Sci. Rev.*, 174, 213–250. doi: 10.1007/s11214-012-9942-9.

Ramirez R.M., Kopparapu R., Zugger M.E., Robinson T.D., Friedmann R., and Kasting J.F. 2014. Warming early Mars with $CO_2$ and $H_2$. *Nat. Geosci.*, 7, 59-63, doi: 10.1038/NGEO2000.

Stern, L. A., Circone, S., Kirby, S. H., Durham, W. B., 2001. Anomalous Preservation of Pure Methane Hydrate at 1 atm, J. Phys. Chem. B 105, 1756-1762.

Tian F., Claire M.W., Haqq-Misra J.D., Smith, M., Crisp D.C., Catling D., Zahnle K., and Kasting J.F. 2010. Photochemical and climate consequences of sulfur outgassing on early Mars. *Earth Planet. Sci. Lett*. 295, 412-418.

Wordsworth R., Forget F., Millour E., Head J.W., Madeleine J.-B., and Charnay B. 2013. Global modelling of the realy martian climate under a denser CO2 atmosphere : Water cycle and ice evolution. Icarus 222, 1-19.